# Coronal magnetic field and the plasma beta determined from radio and multiple satellite observations

Kazumasa Iwai[1*], Kiyoto Shibasaki[1], Satoshi Nozawa[2], Takuya Takahashi[3], Shinpei Sawada[2], Jun Kitagawa[4], Shun Miyawaki[2], and Hirotaka Kashiwagi[5]

1. Nobeyama Solar Radio Observatory, National Astronomical Observatory of Japan, Minamimaki, Nagano 384-1305, Japan; kazumasa.iwai@nao.ac.jp
2. Department of Science, Ibaraki University, Mito, Ibaraki 310-8512, Japan
3. Department of Astronomy, Kyoto University, Sakyo, Kyoto 606-8502, Japan
4. Solar-Terrestrial Environment Laboratory, Nagoya University, Nagoya, Aichi 464-8601, Japan
5. Planetary Plasma and Atmospheric Research Center, Tohoku University, Sendai, Miyagi 980-8578, Japan

**Abstract**

We derived the coronal magnetic field, plasma density, and temperature from the observation of polarization and intensity of radio thermal free-free emission using the Nobeyama Radioheliograph (NoRH) and extreme ultraviolet (EUV) observations. We observed a post-flare loop on the west limb 11 April 2013. The line-of-sight magnetic field was derived from the circularly polarized free-free emission observed by NoRH. The emission measure and temperature were derived from the *Atmospheric Imaging Assembly* (AIA) onboard *Solar Dynamics Observatory* (SDO). The derived temperature was used to estimate the emission measure from the NoRH radio free-free emission observations. The derived density from NoRH was larger than that determined using AIA, which can be explained by the fact that the low temperature plasma is not within the temperature coverage of the AIA filters used in this study. We also discuss the other observation of the post-flare loops by the EUV Imager onboard the Solar Terrestrial Relations Observatory (STEREO), which can be used in future studies to reconstruct the coronal magnetic field strength. The derived plasma parameters and magnetic field were used to derive the plasma beta, which is a ratio between the magnetic pressure and the plasma pressure. The derived plasma beta is about $5.7 \times 10^{-4}$ to $7.6 \times 10^{-4}$ at the loop top region.

## 1. Introduction

In the solar corona, there exist various dynamical phenomena such as flares and coronal mass ejections (CMEs), which are thought to be caused by the interactions between the coronal magnetic field and plasma. Hence, coronal science usually requires a precise measurement of coronal plasma parameters, such as the magnetic field, plasma density, and temperature, which is usually difficult.

Measurement of the coronal magnetic field is especially difficult because of the weak magnetic field strength in hot and turbulent plasma. The coronal magnetic field has been derived via extrapolations of the photospheric magnetic fields using the potential field (Sakurai 1982; Shiota et al. 2008) and linear or nonlinear force-free field approximations (e.g., Inoue et al. 2012). Zeeman and/or Hanle effects in the optical or infrared lines can be used to measure the coronal and chromospheric magnetic fields (Lin et al. 2004; Trujillo Bueno et al. 2005; Hananoka 2005). The coronal magnetic field has also been derived from microwave observations, using gyro-resonance emission (Dulk 1985; Gary and Hurford 1994), polarization reversal by quasi-transverse propagation (Cohen 1960; Ryabov et al. 1999; 2005), and thermal bremsstrahlung (Bogod and Gelfreikh 1980; Grebinskij et al. 2000; Shibasaki et al. 2011; Iwai and Shibasaki 2013).

Radio thermal bremsstrahlung or so-called thermal free-free emission can be used to derive the longitudinal component of the magnetic field (Bogod and Gelfreikh 1980). Iwai and Shibasaki (2013) derived the coronal and chromospheric magnetic fields using a 17 GHz observation from the *Nobeyama Radioheliograph* (NoRH: Nakajima et al. 1994). They found that both coronal and chromospheric components are included in the circularly polarized emission at 17 GHz, and that therefore some additional information or assumptions are required to separate the two components. Only the line-of-sight magnetic field of coronal loops that are outside of the solar disc can be derived without requiring any assumptions because there is no chromosphere in the line-of-sight direction.

The *Solar Terrestrial Relations Observatory* (STEREO) satellites, which are located at different heliocentric angles from the Sun-Earth line, provide different perspectives of the solar corona. For example, coronal loops located over the limb as viewed from the Earth may be observed on the disc by one of the STEREO satellites, permitting us to resolve their horizontal structures. This may be important when we try to extend the measured longitudinal field to the total magnetic field strength.

The intensity of the free-free emission (I) is derived from the emission measure (EM) and temperature (T) of the source region in the optically thin case ($I \propto EM/\sqrt{T}$; Dulk, 1985). The radio free-free emission includes information from all the electrons along the line-of-sight. However, we need information of either the emission measure or temperature to derive the other parameter. The *Atmospheric Imaging Assembly* (AIA: Lemen et al. 2012) onboard the *Solar Dynamics Observatory* (SDO) has several filters centered on extreme ultraviolet (EUV) iron lines. Aschwanden et al. (2013) developed automated temperature and emission measure analysis tools for coronal loops. Using EUV observations from AIA, both the emission measure and the temperature can simultaneously be derived; however, its temperature coverage is limited. By combining the radio intensity observed by NoRH with the temperature and emission measure analysis from the AIA data, a better estimation for the coronal density and temperature should be achieved.

The purpose of this study is to derive the coronal magnetic field, density, temperature, and their combination, the plasma beta, by combining radio observations with EUV observations. The instruments and data sets used in this study are described in Section 2. The data analysis methods and results are provided in Section 3. The reliability of the derived coronal parameters is discussed in Section 4. The paper is summarized and concluded in Section 5.

## 2. Instruments and data sets

NoRH is a radio interferometer dedicated to solar observation. It has 84 antennas, each with a diameter of 80 cm (Nakajima et al. 1994). NoRH observes the full solar disc every 1 s at 17 GHz (both the intensity and circular polarization). NoRH uses the redundancy of the visibilities to calibrate its phase and amplitude. Hence, the visibilities should be calibrated by the redundancy within a timescale shorter than the timescale of the phase variation. In this study, radio images are synthesized every 1 s using the Steer algorithm that was applied for NoRH image synthesis by Koshiishi (2003). Then, images are averaged to reduce statistical noise. Iwai and Shibasaki (2013) investigated the relationship between the averaging time and the lowest detectable signal level of NoRH. They showed that the noise level of the polarization image decreases with increasing averaging time. In this study, 1200 images (20 min) are averaged. The quality of NoRH polarization images occasionally deteriorates because of bad weather. Hence, an event observed during good weather conditions was used.

AIA has several filters in the UV and EUV range and makes observations at a time cadence of 12 s. In this study, six EUV filters (131, 171, 193, 211, 335, and 94 Å) were used to derive the coronal temperature and emission measure. The temperature coverage of these filters is from 0.6 to 16 MK.

STEREO consists of two spacecraft which are in the orbit around the Sun. STEREO_A was located approximately 133.7° from the Sun–Earth line during the observational period of this study. We used EUV images (171, 193, and 304 Å) from the Extreme Ultraviolet Imaging Telescope (EUVI; Wuelser et al. 2004) to detect coronal loop structures around the active region.

3. Data Analysis and Results
3.1 Overview of the observed event
A post-flare loop observed on 2013 April 11 was analyzed in this study. Figure 1 shows the images of the radio intensity (I) and its circularly polarized component (V) observed with NoRH at 17 GHz. The positively polarized component on the western limb is the post-flare loop analyzed in this study, and corresponds to the active region NOAA 11713 (indicated by a white rectangle). The negatively polarized component around N10E12 and the bipolarized component around N21W20 correspond to active regions NOAA 11719 and 11718, respectively. The circularly polarized components from these on-disc polarized sources contained both an optically thin coronal loop component and optically thick chromospheric component (Iwai and Shibasaki, 2013). There were many other polarized regions orthogonally aligned with the polarized active regions, which were all side-lobes of the active regions. In Figure 1, the post-flare region on the west limb is not contaminated by the side-lobes of the other polarized regions.

The three main processes that produce microwave solar radio emission are free–free emission, gyro-resonance emission, and gyro-synchrotron emission. Gyro-resonance emission is emitted from sunspots at lower harmonics (second or third) of their local gyro-frequency (Shibasaki et al. 1994; Nindos et al. 2000; Vourlidas et al. 2006). The third harmonic of the gyro-frequency at 2000 G is about 17 GHz. We checked the location of the sunspot region observed a few days before and calculated the solar rotation. It is confirmed that the sunspot region NOAA 11713 had already moved completely behind the solar limb at the time of observation. Hence, there was no gyro-resonance emission component from the observed active region.

Gyro-synchrotron emission is emitted from non-thermal electrons during flares. Figure 2 (top panel) shows the time variation of the solar soft X-ray flux observed with the Geostationary Operational Environmental Satellites (GOES). There was a C3.9 flare on April 10 23:30 UT and a C2.2 flare on April 11 0:14 from the NOAA 11713 region. Figure 2 (bottom panel) shows the time variation of the EUV flux of the limb active region (indicated by a white rectangle in Figure 3(c)) observed with EUVI at 193 Å. There were two flux enhancements that corresponded to the two X-ray flares. There was no other enhancement, indicating that there was no "hidden" flare behind the limb. Therefore, the quiet period between 2:35 and 2:55 UT was selected for this study to avoid contamination by gyro-synchrotron emission.

From these observational characteristics, the radio emission analyzed here, which was observed from the post-flare region of NOAA 11713, is assumed to be pure thermal free–free emission.

**3.2 Locations of the polarized radio emission and the coronal loops**
Figure 4(a) shows an EUV image at 304 Å observed with AIA. The white and red contours show the radio intensity and the degree of circular polarization at 17 GHz, respectively. The peak of the radio intensity occurred in the bright region at 304 Å. On the other hand, the degree of circular polarization reached its peak at a higher altitude above the solar limb than the radio intensity (indicated by the dashed white rectangle in Figure 4(a); hereafter called the "loop top (LT) region"). Figure 4(b) shows an EUV image at 171 Å observed with AIA. There were many coronal loop structures around the most circularly polarized region, and the tops of some of the coronal loops corresponded to the peak region of the radio circular polarization. This relationship is discussed in detail in Section 4.

The dynamic range of the brightness temperature image is defined by the ratio between the peak brightness and the standard deviation of the sky region. The peak brightness temperature in Figure 1 (left panel) is 40300 K at the core of AR 11713. The standard deviation of the brightness temperature of the sky region surrounded by the red rectangle Figure 1 is 110 K. Hence, aforementioned ratio is about 362:1. The average brightness temperature of LT region in Figure 4(a) is about 9000 K.

The standard deviation of the polarization components in the quiet region of the solar disc should ideally be 0. This value can serve as a proxy for the polarization accuracy of the instrument (Iwai and Shibasaki, 2013). The standard deviation was about 11 K after averaging over 20 min in the region indicated by the dashed white rectangle in Figure 1 (right panel). We define the five-sigma value (55 K) as the minimum detectable signal level. Hence, the minimum detectable level for the degree of polarization is about 0.5% with an average intensity of 10000 K. The observed degree of polarization in the LT region is 2.7%.

### 3.3 Emission measure and temperature analysis from radio and EUV observations

The column emission measure (EM) and temperature (T) of the coronal plasma can be derived from the EUV flux (hereafter the EM means the column emission measure in this paper). Automated temperature and emission measure analysis tools for AIA data (Aschwanden et al. 2013) were used in this study. The EUV flux observed with an AIA filter on pixel [x, y] $(= F_\lambda(x, y))$ is defined as

$$F_\lambda(x, y) = \int \frac{dEM(T,x,y)}{dT} R_\lambda(T) dT ,  \quad (1)$$

where $R_\lambda(T)$ is the response function of the filter and $dEM(T,x,y)/dT$ is the deferential emission measure (DEM). We used the latest response function of AIA (version 6) with CHIANTI version 7.1.3. The empirical correction of the missing emission lines to 94 and 131 Å channels is performed because it is suggested that there is a deficiency of the CHIANTI database in the 50–170 Å wavelength range (Boerner et al 2014).

The DEM of each pixel is approximated as a Gaussian function,

$$\frac{dEM(T,x,y)}{dT} = EM_P(x, y) \exp\left(-\frac{[\log(T) - \log(T_P(x,y))]^2}{2\sigma_T^2(x,y)}\right) , \quad (2)$$

where $EM_P(x, y)$, $T_P(x, y)$, and $\sigma_T(x, y)$ are the peak emission measure, peak temperature, and width of the Gaussian temperature of each pixel, respectively. Fitting the observed flux using Equation 2 and the response functions of AIA filters produce $EM_P(x,y)$, $T_P(x,y)$, and $\sigma_T(x,y)$ for each pixel. Figures 4(c) and 3(d) show the derived peak emission measure and peak temperature, respectively. The spatial distribution of the emission measure derived from the AIA observations corresponded well with the 17 GHz radio emission region. The region with the highest level of 17 GHz emission had a higher temperature than the ambient corona.

The average DEM of the loop top region was derived by averaging the single pixel DEMs in that region. Figure 4(f) shows the average DEM in the LT region. The averaged column emission measure of this region is $1.3 \times 10^{25} (\text{cm}^{-5})$, which was derived by integrating the DEM with respect to the temperature.

The emission measure and temperature of the coronal plasma was also derived from the radio thermal free-free emission. The opacity (τ) or absorption coefficient (κ) of the thermal free-free emission is derived as follows (Dulk 1985):

$$\tau = \int \kappa dl \propto g_{ff} T^{-\frac{3}{2}} \nu^{-2} \int n_e n_i dl \propto g_{ff} T^{-\frac{3}{2}} \nu^{-2} EM , \qquad (3)$$

where ν and $g_{ff}$ are the radio frequency and Gaunt factor, respectively. $n_e$ and $n_i$ are the electron and ion density, respectively, and a fully ionized electroneutral atmosphere is assumed. The Gaunt factor is a function of frequency and temperature (Dulk 1985). It depends very weakly on the temperature and can be recognized as a constant in this study.

The brightness temperature of the corona ($T_B$) is approximated as $T(1 - e^{-\tau})$, where τ is the opacity of the corona. The typical temperature of the LT region is about 3 MK in Figure 4(d), and the brightness temperature of this region is about 9000 K in Figure 4(a). The opacity of the LT region is derived to be about 0.003; hence, the optically thin approximation can be applied. In the optically thin region, the brightness temperature of the free-free emission is given by the coronal electron temperature and coronal opacity,

$$T_B \approx T\tau \propto \nu^{-2} \frac{EM}{\sqrt{T}} \propto \lambda^2 \frac{EM}{\sqrt{T}} , \qquad (4)$$

where λ is the wavelength. In this study, the peak temperature derived from the AIA observation shown in Figure 4(d) was used to derive the emission measure of the 17 GHz observation. Figure 4(e) shows the emission measure derived from the brightness temperature at 17 GHz and the temperature of the AIA observation. The spatial distribution of the emission measure derived from the NoRH data corresponded well with that of the peak emission measure from the AIA data. The average emission measure of the LT region was $4.2 \times 10^{25} (\text{cm}^{-5})$.

### 3.4 Magnetic field analysis using the radio and STEREO observations

The opacity of the free–free emission is defined by Equation 3, given in the previous section. However, this is an averaged opacity of the ordinary (O) and extraordinary (X) modes of the free–free emission, and these two modes have different optical depths in a magnetized plasma. This means that a circularly polarized component is generated. The actual opacities of the O- and X-modes of the free-free emission ($\tau_{\{O,X\}}$) are defined as follows:

$$\tau_{\{O,X\}} \propto (\nu \pm \nu_B |\cos \alpha|)^{-2} \, T^{-\frac{3}{2}} \, EM \, , \qquad (5)$$

where $\nu_B$ is the local gyro-frequency, and $\alpha$ is the angle between the local magnetic field and the line-of-sight direction. The circularly polarized component of the free-free emission is inverted to obtain the longitudinal component of the magnetic field $B_{los}$ as follows (Bogod and Gelfreikh 1980):

$$B_{los}[G] = \frac{10700}{n\lambda[cm]} \frac{V}{I}$$

$$n = \frac{d(\log I)}{d(\log \lambda)} \, , \qquad (6)$$

where $I$ is the brightness temperature ($I = T_B$), $V$ is the brightness temperature of the circularly polarized component, and $n$ is the power-law spectral index of the brightness temperature.

We can derive the spectral index in the optically thin case using Equations 4 and 6 as follows:

$$n = \frac{d(\log I)}{d(\log \lambda)} \approx \frac{d}{d(\log \lambda)} \log \left( C \lambda^2 \frac{EM}{\sqrt{T}} \right) = 2 \, , \qquad (7)$$

where C is a constant. Hence, the optically thin assumption in the coronal environment gives a spectral index of about 2. In fact, the spectral index of the microwave free-free emission in the corona is typically 2 (e.g. Kundu 1965). The average degree of circular polarization of the LT region is about 2.7%, from which the longitudinal component of the magnetic field was calculated to be about 84G.

The magnetic field vector has three components (line-of-sight, east-west, and north-south). The magnetic field strength is given by

$$|B|^2 = (B_{los})^2 + (B_y)^2 + (B_z)^2 ,  \qquad (8)$$

where |B| is the magnetic field strength and $B_y$ and $B_z$ are the east-west and north-south components of the magnetic field, respectively. We assume that the magnetic field at the loop top is parallel to the solar surface. Hence, the magnetic field vector at the loop top is in the plane parallel to the tangent plane of the surface. At the extreme limb, the line-of-sight vector is in this plane ($B_y = 0$). We define the angle between line-of-sight and coronal loops as θ. Hence, the line-of-sight component is given by

$$B_{los} = |B| \cos \theta . \qquad (9)$$

Figure 3(c) shows the EUV image at 171 Å observed with EUVI onboard STEREO_A. The region enclosed by the white rectangle corresponds to the limb active region examined in this study. Figure 3(b) shows a close-up of the active region that is enclosed by the white rectangle in Figure 3(c). There were several characteristic loop structures that corresponded to the AIA image. For example, there was a large loop structure that crossed over the center of the active region from north to south. Two bright footprints of the loop structures (indicated by "FP1" and "FP2" in Figure 3(b)) correspond to FP1 and FP2 in Figure 4(b). Middle of the FP1 and FP2 is considered to be the most circularly polarized region of 17 GHz. Figure 3(d) shows a high-pass filtered image of Figure 3(b). There were several loops along the line-of-sight of the LT region (indicated by the green line in Figure 3(d)). The magnetic field derived from the radio observation is the emission measure weighted magnetic field. Hence, the brighter loops can have greater influence on the magnetic field. We focus on the loops indicated by the blue line in Figure 3(d). These loops were located around the extreme limb from the Earth's view, and they were brighter than surrounding loops. Hence, it is suggested that these structures are the origin of the main component of the circularly polarized radio emission of the LT region. The inclination of this loop is about 10° to the line-of-sight direction from the Earth.

STEREO_A was located 133° from the Sun-Earth line. Hence, the active region was observed in the eastern hemisphere by STEREO_A (see Figure 3(c)). That means the tilt angle above is not the actual angle but the apparent angle. The two footpoints of the loops are separated about 100 arc seconds in an east-west direction (that is the line-of-sight direction from the Earth) and 60 arc seconds in a north-south direction. If we assume the loops were the simple arc structure that connected the two footpoints, the tilt angle from the line-of-sight direction was about 31° as shown in the red line in Figure 3(d). The actual separation between the two footpoints in an east-west direction should be larger than it was looked from STEREO_A because the active region was not in the disc center in the STEREO_A view. That means the actual tilt angle should be smaller than 31°. Unfortunately, the triangulation method is not feasible, because the separation angle between STEREO and SDO is excessively large. Therefore, we cannot derive the actual tilt angle in this study. However, the tilt angle of the main loop structures may be in the range between 10 and 31°. Therefore, the magnetic field strength of this region was calculated to be from 85 to 98G.

In Figure 3(d), the loop structures at the LT region have a width of several arcsec (3000 ~ 4000 km). We estimated the line-of-sight depth of the LT region to be 3500 km, and assumed the plasma distributed uniformly within the line-of-sight depth. This effect is discussed in detail in Section 4. The plasma density measured at the LT region was $3.5 \times 10^8$ (cm$^{-3}$) from NoRH and $2.0 \times 10^8$ (cm$^{-3}$) from AIA. Hence, the density derived from the radio observation is about 43% larger than that derived from the EUV observation.

Using the derived magnetic field, plasma density, and temperature of the LT region, we calculate the plasma beta (plasma pressure/magnetic pressure) to be $5.7 - 7.6 \times 10^{-4}$. The derived plasma parameters of the LT region are summarized in Table 1.

## 4. Discussion
### 4.1 Difference of the emission measures derived from the radio and EUV observations

This study derived the emission measures using two ways: from the radio free-free emission and from the EUV line emission. The emission measures at the LT region derived from the radio observation is larger than that derived from the EUV observation.

The free-free emission is emitted from all the electrons. On the other hand, the EUV observation derives the plasma density only within the temperature range over which the filters are sensitive. Hence, the DEM derived from the EUV observation is considered as a lower limit and it is usually smaller than that derived from the radio observation (e.g. Alissandrakis et al 2013).

For this study, a flare quiet period was selected using GOES and STEREO_A data (see Section 3.1). Hence, contributions from high temperature plasma (T > 16 MK) were negligible. On the other hand, there was emission at 304 Å from the LT region, as shown in Figure 4(a). This emission suggests that there was low temperature plasma that was not included in the density derived from the analysis of the AIA data (0.6 MK < T < 16 MK). In addition, this study used a single Gaussian model to derive the DEM. Hannah and Kontar (2012) estimated the accuracy of the single Gaussian inversion using SDO/AIA data. They found that the typical range for the temperature errors is $\Delta \log T \approx 0.1 - 0.5$, which is dependent on the observed signal-to-noise ratio. A similar error range is likely to be included in our results. It is also suggested that the current calibration of AIA is not perfect (Boerner et al. 2014). Although the evaluation of the AIA calibration is beyond of the purpose of this study, it should be noted that the comparison with radio and EUV observations can be a cross-calibration for both observations.

The emission measures derived from the radio observation may also contain errors. The peak temperature derived from the EUV observations was used to derive the emission measure in Equation 4. However, the actual temperature that affects the radio free-free emission is an "emission measure weighted" temperature that contains all electrons across all temperature ranges. If the DEM is a symmetric Gaussian distribution whose center is the peak temperature, the peak temperature can be used as the typical temperature of the free-free emission. However, the observed temperature shown in Figure 4(f) is not symmetric. Moreover, low temperature plasma also exists outside of the temperature range of Figure 4(f), as noted above. However, the effect of the temperature error would be relatively small because the brightness temperature varies with the emission measure and the inverse of the square root of the temperature in Equation 4.

For these reasons, it is suggested that the emission measures derived from the two independent methods are explained consistently, and that the primary reason why the EUV emission measure is estimated to be smaller than the emission measure derived from the radio observations is the existence of low temperature plasma.

**4.2 Location of the polarized emission and the coronal loop structure**

In Figure 4, the most polarized region (the LT region) is located about 50 arc seconds (= 35,000 km) above the solar surface. In general, the magnetic field strength should be stronger in lower altitude regions. However, the degree of circular polarization is also affected by the line-of-sight angle of the magnetic field and the opacity of the free-free emission.

Figure 3(a) shows the 304 Å image observed with STEREO_A. There were bright loops around the center of the active region (labeled "AR"). These seemed to correspond to the bright region of 17 GHz emission (also labeled "AR") in Figure 4(a). There were many small scale loop structures with their earthward and anti-earthward footpoints. The spatial resolution of NoRH is about 10 arc seconds, which is insufficient to resolve an individual loop element. Hence, there is a possibility that the superposition of the polarized emission from various loop directions reduced the observed circularly polarized emission of the AR region.

**4.3 Evaluation of the plasma parameters at the loop top**

From the Baumbach-Allen electron density model (Allen 1947), the typical density above 35,000 km from the solar surface is $2.5 \times 10^8$ (cm$^{-3}$). A 3- to 10-fold Baumbach-Allen electron density model is usually used for active regions. The density derived from the NoRH observation was $3.5 \times 10^8$ (cm$^{-3}$). Hence, the derived density is of the same order as the typical coronal density.

In this study, we estimated the line-of-sight depth of the LT region to be 3500 km from the loop width by the STEREO observation. Hence, it is assumed that the line-of-sight spatial scale was the same as the loop width, and that plasma distributed uniformly within the line-of-sight depth. However, the actual plasma may distribute non-uniformly (the filling factor along the line-of-sight direction was not constant). Moreover, the structures of radio and EUV emission regions are different (see Figure 4). That means the radio and EUV emission measures required the different filling factors. However, it is difficult to estimate the actual filling factor of the radio emission and further investigations are required. We note that tomographic observations using SDO and STEREO will enable a more accurate estimation.

From Dulk and Mclean (1978), the typical magnetic field strength above 35,000 km from the solar surface is about 44 G; this value is of the same order as the derived magnetic field (85 - 98 G). In this study, the time of the flare occurrence that generated the gyro-synchrotron emission was strictly eliminated using GOES and STEREO_A. Contamination by gyro-resonance emission is also negligible because the sunspot area of the active region was behind the solar limb as viewed from the Earth. Hence, the observed emission was pure thermal free-free emission. The observed degree of circular polarization was about 2.7%, which is larger than the detection limit of NoRH (about 0.5%) estimated by Iwai and Shibasaki (2013). In this study, the tilt angle of the loop structure was derived using the STEREO observation. Although the derived tilt angle is more accurate than those estimated from Earth-based detection, the triangulation method is not feasible in this study. Hence, the derived inclination angle has errors. This is a limitation of this study, and further investigation for determining the loop geometry will be required in the future.

The plasma beta value of the LT region was about $5.7 - 7.6 \times 10^{-4}$. The typical beta value of the lower corona is between $10^{-3} \sim 10^{-1}$ (Gary 2001). Hence, the derived beta in LT region was a little bit smaller than the typical value. The derived density was the averaged density in the LT region, and the derived magnetic field was the emission measure weighted magnetic field along the line-of-sight direction. The bulk of the plasma particles was concentrated in the coronal loops. Hence, the plasma density inside the coronal loop may have been larger than the average density of the LT region, while the magnetic field of the coronal loops did not vary largely from the emission measure weighted magnetic field because the coronal loops contained the majority of the emission measure. Therefore, the actual plasma beta of the LT region may be larger than the derived beta value.

**5 Summary and Conclusion**

We have investigated the magnetic field, density, and temperature of solar corona using radio and EUV observations. Radio free-free emission observed by NoRH was combined with multiple line-of-sight EUV observations from STEREO and SDO, which enabled more accurate estimations of the coronal parameters than have ever previously been derived. The observational results are summarized as follows.

- In this study, the density was derived using two methods. The emission measure and temperatures are derived from AIA observations using six filters. The derived temperature was used to estimate the emission measure of the radio free-free emission observed by NoRH. The density derived from the radio observation was 43% larger than that derived from the AIA observation. The difference between the two densities can be explained by the presence of low temperature plasma that was out of the temperature coverage of the EUV filters used in this study.
- The line-of-sight magnetic field was derived from the circularly polarized emission observed by NoRH. This could be extended to a magnetic field strength using STEREO observations. The strongest line-of-sight magnetic field was observed at 35,000 km above the solar surface.
- The plasma density of the coronal loops may have been larger than the derived average density, while the magnetic field of the coronal loops did not vary significantly from the derived emission measure weighted magnetic field. That may cause the plasma beta to be smaller than the typical value.

In this study we derived the coronal magnetic field by combining radio and EUV observations. In addition, in this study we derived the plasma density from radio and EUV observations, and calculated the plasma beta. Although these values are spatially averaged or emission measure weighted, the better estimations were provided by combining different observational data. For the next step, radio observations with higher spatial resolution will be required to resolve the coronal loop structure and derive its intrinsic plasma parameters. This requirement will be met using recent and future radio interferometers such as the SSRT (Lesovoi et al. 2012), CSRH (Yan et al. 2009), and FASR (Bastian 2004).

**Authors' contributions**
KI leaded and managed this study and drafted the manuscript. KS summarized the scientific background and helped to draft the manuscript. SN, TT, SS, JK, SM, and HK carried out the data analysis. All authors read and approved the final manuscript.

**Acknowledgments**
This study is based on results from the Coordinated Data Analysis Workshop 2013 (CDAW2013) organized by the Nobeyama Solar Radio Observatory, NAOJ, and the Solar-Terrestrial Environment Laboratory, Nagoya University. SDO data are courtesy of NASA/SDO and the AIA science teams. The SECCHI data are produced by an international consortium of Naval Research Laboratory, Lockheed Martin Solar and Astrophysics Lab and NASA Goddard Space Flight Center (USA), Rutherford Appleton

Laboratory and University of Birmingham (UK), Max-Planck-Institut für Sonnensystemforschung (Germany), Centre Spatiale de Liege (Belgium), Institut d'Optique Théorique et Appliquée, and Institut d'Astrophysique Spatiale (France).

**Figures**

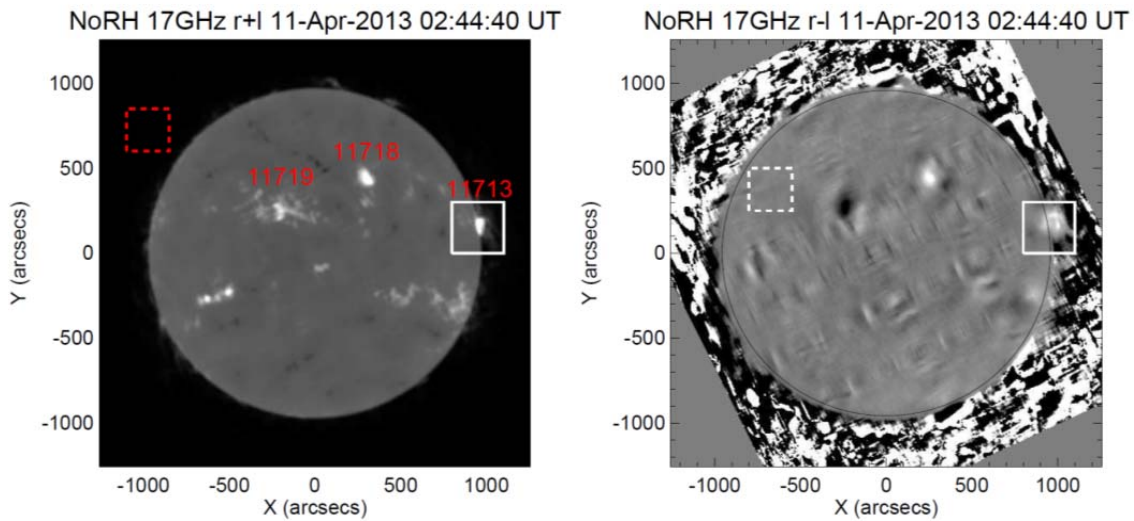

**Figure 1**: Radio intensity and circularly polarized component observed with NoRH (Left) Intensity and (right) circularly polarized component of the radio emission at 17 GHz observed with NoRH on 2013 April 11 02:44:40 UT.

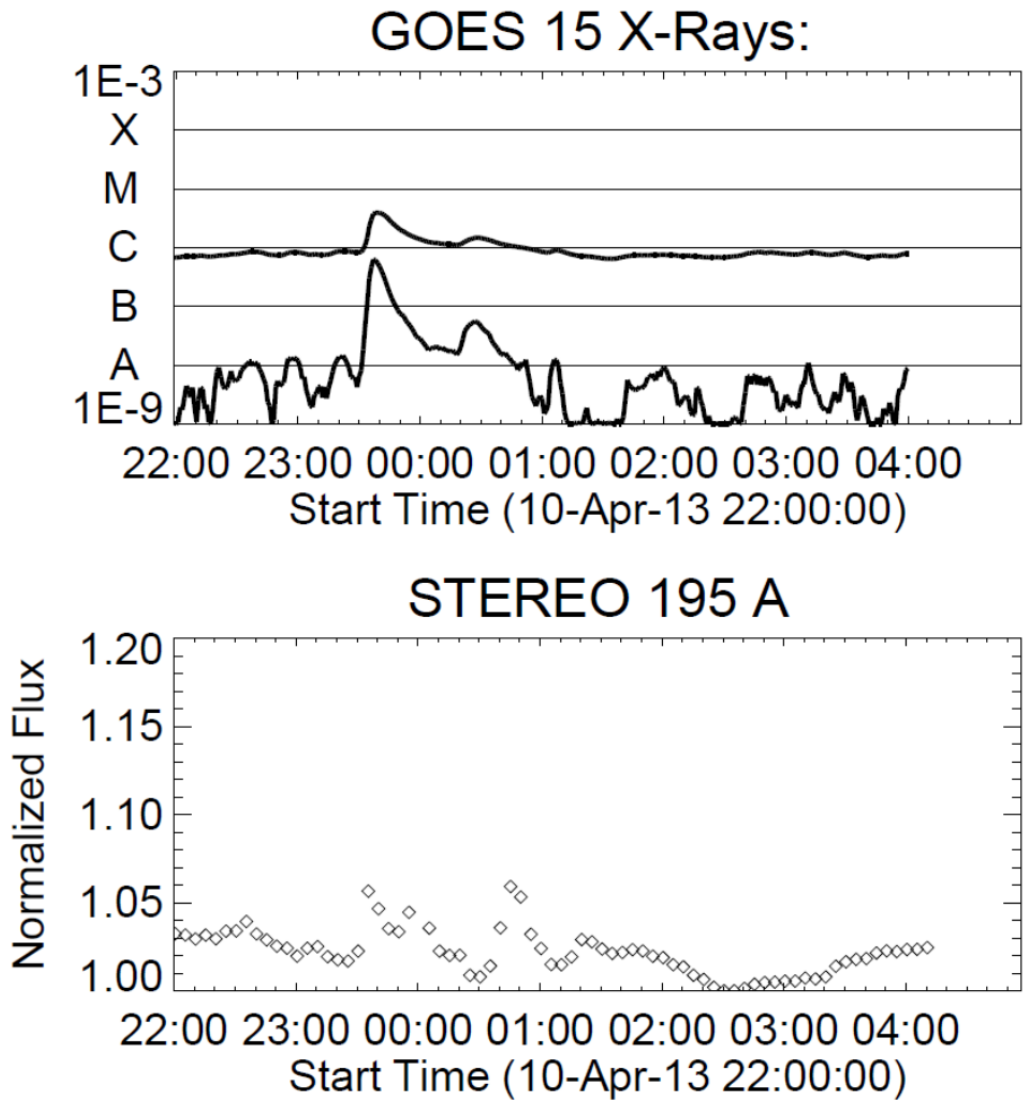

**Figure 2**: Time variation of the total soft X-ray and EUV flux of the active region
(Top) Time variation of the total soft X-ray flux of the Sun observed with GOES 15. (Bottom) Time variation of the EUV flux at 195 Å observed with STEREO_A. The flux plotted is the average flux over the active region marked by the white rectangle in Figure 3d.

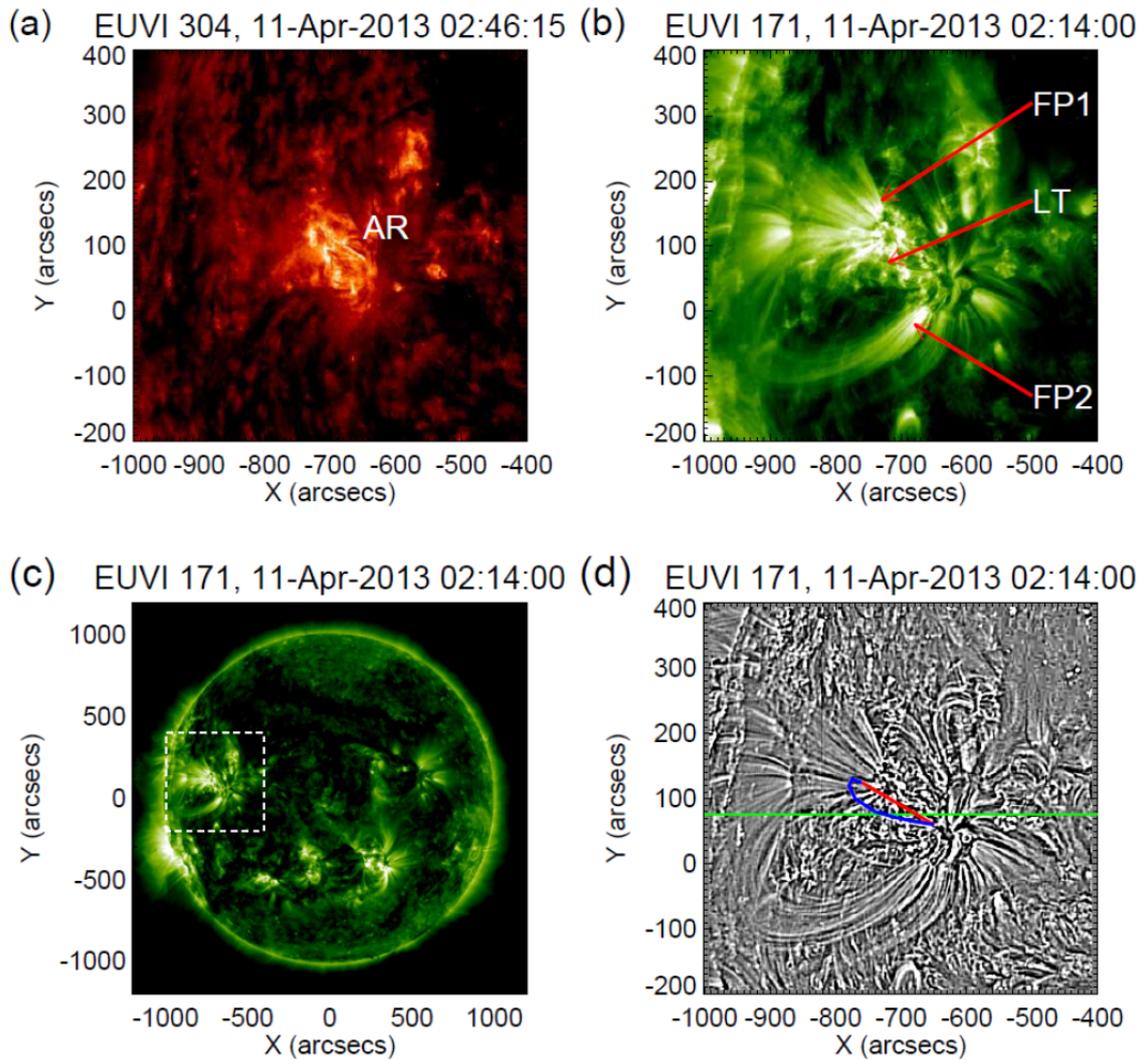

**Figure 3:**

(a) EUV image of the active region enclosed by the white rectangle in panel (c) at 304 Å and (b) at 171 Å, observed by AIA. (c) EUV image of the entire solar disc at 171 Å, observed by AIA. (d) High-pass filtered image of panel (b).

Results of the radio and EUV observations of the active region

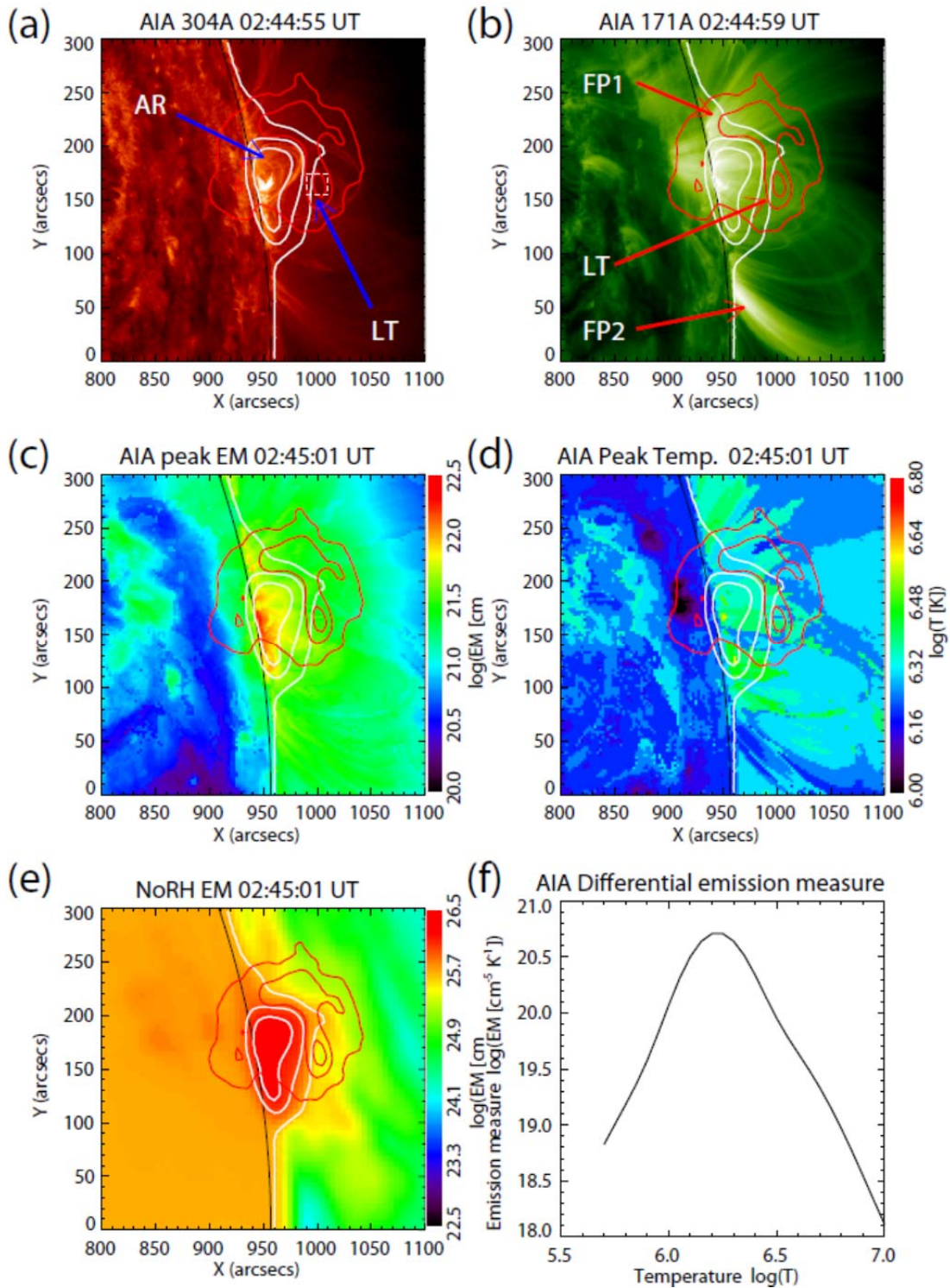

**Figure 4:** EUV observation from STEREO_A
(a) EUV image at 304 Å observed with AIA. (b) EUV image at 171 Å observed with AIA. (c) Peak emission measure derived from AIA observations. (d) Peak temperature derived from AIA observations. (e) Emission measure derived from the brightness temperature at 17 GHz and the temperature from AIA using the optically thin assumption. Note that optically thick emission, corresponding to the brightness temperature of the τ = 1 layer, was observed in the on-disc region. Hence, the emission measure for the region inside the solar disc is incorrect. (f) Total deferential emission measure (DEM) in the region enclosed by the dashed white rectangle in panel (a). White contours in panel (a) to (e) show the radio intensity at 17 GHz (levels = 10000, 20000, 30000 K). Red contours in panel (a) to (e) show the degree of circular polarization (levels = 1.0, 2.0, 3.0 %) at 17 GHz.

**Table 1:** Plasma parameters of the LT region
Plasma parameters of the LT region derived from NoRH, AIA, and STEREO observations

| Parameters | Values |
|---|---|
| 17 GHz V/I (%) | 2.7 |
| $B_l$ (G) | 84 |
| Tilt angle (degree) | 10 - 31 |
| $B_{abs}$ (G) | 85 - 98 |
| AIA Temp (Log K) | 6.4 |
| AIA $n_e$ (cm$^{-3}$) | $2.0 \times 10^8$ |
| 17 GHz $n_e$ (cm$^{-3}$) | $3.5 \times 10^8$ |
| Plasma beta ($\times 10^{-4}$) | $5.7 - 7.6$ |